\documentclass{article}
\usepackage{arxiv}

\usepackage[utf8]{inputenc} 
\usepackage[T1]{fontenc}    
\usepackage{hyperref}       
\usepackage{url}            
\usepackage{booktabs}       
\usepackage{amsfonts}       
\usepackage{nicefrac}       
\usepackage{microtype}      
\usepackage{lipsum}

\usepackage{float}
\usepackage{booktabs} 
\usepackage{algorithm}
\usepackage{amsmath}
\usepackage{amssymb}
\usepackage{graphicx}
\usepackage[noend]{algpseudocode}
\usepackage{siunitx}
\usepackage{dcolumn}
\algnewcommand\algorithmicforeach{\textbf{for each}}
\algdef{S}[FOR]{ForEach}[1]{\algorithmicforeach\ #1\ \algorithmicdo}
\def\BState{\State\hskip-\ALG@thistlm}
\newcolumntype{d}[1]{D{.}{.}{#1}}
\newcommand{\passthrough}[1]{#1}
\newcommand{\commentBodo}[1]{{\passthrough{\color{blue}[BB thinks: {\emph{#1}}]}}}

\title{Scalable Methods for Calculating Term Co-Occurrence Frequencies}

\author{
    Bodo Billerbeck\\
    Microsoft and\\University of Melbourne\\
    \url{bodob@microsoft.com}\\
    \And
    Justin Zobel\\
    University of Melbourne\\
    \url{jz@unimelb.edu.au}\\
    \And
    Nicholas Lester\\
    Microsoft\\
    \url{nlester@microsoft.com}\\
    \And
    Nick Craswell\\
    Microsoft\\
    \url{nickcr@microsoft.com}\\
}

\begin{document}
\maketitle



\begin{abstract}
Search techniques make use of elementary information such as term frequencies and document lengths in computation of similarity weighting.
They can also exploit richer statistics, in particular the number of documents in which any two terms co-occur.
In this paper we propose alternative methods for computing this statistic, a challenging task because the number of distinct pairs of terms is vast -- around 100,000 in a typical 1000-word news article, for example.
In contrast, we do not employ approximation algorithms, as we want to be able to find exact counts.
We explore their efficiency, finding that a na\"ive approach based on a dictionary is indeed very slow, while methods based on a combination of inverted indexes and linear scanning provide both massive speed-ups and better observed asymptotic behaviour.
Our careful implementation shows that, with our novel \textsc{list-pairs} approach it is possible to process over several hundred thousand documents per hour.
\end{abstract}


\section{Introduction}\label{section:introduction}

A document can be viewed as a bag of terms, on which statistics can be calculated in order to yield similarity scores.
In addition to elementary statistics such as word frequency, some methods make use of the number of documents in which any pair of terms co-occurs.
These terms need not be adjacent, but can be anywhere in the document; or they may be confined to a window or controlled by a distance parameter.
Methods that make use of this kind of statistic include PMI \cite{church1990word}, PPMI \cite{turney2010frequency,aji2012summarization}, LSI \cite{landauer1998introduction}, and deep-learning techniques such as GloVe \cite{pennington2014glove} and Word2Vec \cite{mikolov2013distributed}.

Methods such as GloVe and Word2Vec use heuristics to predict co-occurring terms to achieve scalability while others, such as PMI, LSI and PPMI, use statistics to predict the most probable co-occurrences.
%
Typically these methods use heuristics to achieve scalability.
For example, they might use a small window of terms and a limited memory size for a scalable co-occurrence count \cite{goyal2011approximate}, or
choose a seed term as the starting point of counting and examine the seed's co-occurring terms~\cite{buzydlowski2002coanalysis}.
These approaches are reported to be scalable but inaccurate \cite{turney2003measuring}.
While use of windows can have the benefit of counting only term pairs that appear in a stricter context, this may not always be desirable; for example, Mitra and Craswell \cite{mitra2017neural} show that using the whole document as context can lead to distinct, desirable outcomes.
As another contrast, this task is distinct from pattern mining; the goal is not just to find common co-occurrences, or anomalous co-occurrences, but to have statistics for all of them.

Within a document (or window), the number of co-occurring pairs is quadratic in document length, and a collection of a million documents can easily contain ten billion words pairs -- as is the case for our test collection.
Since the number of pairs is such that a complete dictionary of pairs cannot be held in reasonable volumes of memory, in broad terms one of two approaches must be used.

One approach is to proceed in document order, flushing information to disk once memory is saturated, followed by a merge of the blocks of flushed information at the end.
Each document is then inspected once only, but the cost of merging, and size of intermediate files, can be significant.
The other approach is (loosely speaking) to proceed in term order, so that all pairs involving a particular term are identified at the same time, but with some volume of repeated inspection of documents. 
In principle such approaches should have lower asymptotic cost, since the number of repeats is bounded by document lengths (which do not grow with collection size), but it is plausible that constant factors will mean that in practice such approaches are expensive.

We are not aware of previous explorations of this question in the literature.
While there are numerous methods that make use of such counts \cite{buzydlowski2002coanalysis,goyal2011approximate,turney2003measuring}, how best to create them is not clear.
Previous attempts to obtain co-occurrence counts include ``Google counts''~\cite{matsuo2007robust}, whereby search engines (such as \url{www.google.com}) are queried for `termA AND termB' and the displayed result count is taken as best evidence.
There have been previous attempts to count term co-occurrences by using high performance distributed systems in order to cope with the significant amount of required resources (memory and time) \cite{pantel2009webdistribution}, but such systems are often not available and cost-effective to a standard user. Therefore our goal is to overcome this issue by introducing new methods that can simply work on a standard machine that is available to an average user.

We have developed several methods for co-occurrence counting, including a na\"ive baseline and methods based on combinations of inverted lists and scanning.
Our experiments on a collection of over 1,000,000 documents show that a method based on processing in term order is the fastest, and appears to have the best asymptotic behaviour.
With over 10 billion distinct term pairs in the collection, we observed factor-of-10 to factor-of-100 differences in efficiencies between the methods; the fastest and most-memory efficient was able to process the collection in a few hours on a modest machine.
These times could of course be reduced through parallel computation, but we anticipate that the relative times would be the same, while noting challenges in parallelizing these order-dependent algorithms

\section{Methods}\label{section:methods}

Our proposed methods make use of some common structures.

One of these is the documents themselves.
If they are to be inspected multiple times, preprocessing into a simplified form can yield significant efficiency gains.
We replace each distinct term by an ID, which is an ordinal 32-bit integer; we only record each ID once per document, and sort the term IDs into ascending order.
Ascending term IDs are assigned to unique terms as these are encountered in the collection.

We build an elementary inverted index for the collection, which is a structure that stores, for each term, a list of the identifiers of the documents containing that term \cite{zobel2006inverted}.

We note that for all methods we are careful to only consider distinct pairs, that is, we assume a lexicographic ordering on the terms and only count a pair A-B where A is less than B.
We refer to the least term of the pair as a primary key and the greater as a secondary key.

In the following we describe our implemented methods:

\paragraph{\textsc{na\"ive}}
We call our baseline the \textsc{na\"ive} method, as it is arguably the most straightforward algorithm:
In this method, for each document, all distinct combinations of term pairs are found, and their counts are incremented.
As not all term pairs can be kept in memory simultaneously, after a certain number of pairs (a controllable parameter) have been accumulated we write these out to disk in a temporary file (a primary key followed by multiple tuples of secondary keys and counts -- we use this output format for the final output, too, and for all our methods).
While typically these would be later on merged in an external process, we can accelerate this (to a limited extent) by keeping in-memory file offsets to the primary keys.
As our results show, however, the costs remain substantial.
In our experiments, we chose -- somewhat arbitrarily -- to flush the memory whenever 100M distinct pairs have been accumulated.

\paragraph{\textsc{list-pairs}}
An alternative to accumulating term pairs document by document, as each pair is encountered, is to accumulate these in pairs directly.
Simplistically,
one could take the two first terms encountered and stream through the document collection to find all occurrences, write those counts to disk, and then repeat for the next pair of terms.
Instead we build a simple inverted index in a first pass over the collection.
We then consider all possible (ordered) term pairs and calculate the intersection of their lists, which yields the final count that is then immediately committed to disk.
The advantage of this approach is that -- apart from the initial pass over the collection -- each pair is considered exactly once only and only a single count has to be accumulated in memory.
The disadvantage is that the approach is quadratic in the number of distinct terms and in the number of term occurrences, and the space of pairs is extremely sparse; the vast majority of the calculated intersections will be empty.

\paragraph{\textsc{list-blocks}}
We can improve on the previous method by aggregating the inverted lists into $b$ blocks of up to $k$ lists each.
Within each block, the postings are organised no longer by term, but by the documents they originally appeared in; that is, the blocks contain smaller versions of (some of) the original documents, but each of those contain only a subsection of the vocabulary.
Once the blocks are constructed they are compared pairwise with each other.
This is done by holding one of the blocks constant as the ``outer'', and pairing this up with all remaining blocks in the ``inner'' position.
For the duration of the block being in the ``outer'' position, the constituent terms are considered as the primary keys, while all other terms (contained in ``inner'' blocks) are designated as secondary keys.
While pairing blocks, the set of matching pairs of documents are considered.
Each of these pairs are processed by generating all possible combinations of primary keys with secondary keys, and adding a count of one to a corresponding accumulator.
This process is repeated with all other inner blocks.
Finally all possible ordered pairs of terms within each document of the outer block must be generated and counts added to the accumulators, before the existing accumulators can be written to disk.
Once completed, all the co-occurrences with terms in the outer block have been counted and need no longer be considered; the outer block can therefore be discarded.
In order to maintain the ordering all primary keys being smaller than secondary keys, the blocks must be organised such that the smallest keys are in the first blocks, followed by the next smallest keys, etc.
Blocks are then processed in order of the keys they contain.
Thus only $b*(b+1)/2$ blocks need to be processed: one pass ($b$) of all blocks to do the inner join of the lists contained within each block (arguably, this could be part of the following nested structure), and then a nested loop to get all unordered pairings of blocks.

A natural block size is the square root of the number of inverted lists (the vocabulary size), which is then also roughly equal to the number of blocks (that is, $b \approxeq k$).
Experiments, not reported here, confirm this to be a good choice.

\paragraph{\textsc{list-scan}}
The approach \textsc{list-scan} requires the construction of a forward index (a list of all unique terms appearing in a particular document; in other words: a simplified version of the document itself) alongside an inverted index \cite{heinz2003efficient}.
To accumulate counts, the inverted list (comprising the primary key) of each term is scanned in turn and each of the referenced forward documents are loaded.
An accumulator table keeps a count for each secondary key encountered.
Once all documents that the current primary key is contained in have been scanned, the resulting counts can be written out and the next inverted list is scanned.
This method means that each document is inspected at most once for each term it contains, and so is in principle linear assuming a bound on document size, but the constant factors may be high.

\paragraph{\textsc{multi-scan}}
In our final method we construct the forward documents and then keep a number (a controllable parameter) of accumulator tables in memory.
The forward documents are scanned in turn and the first $a$ terms encountered are considered the primary keys for the current scan and are allocated accumulator tables.
For all documents that contain one of the primary keys, all terms with a higher term ID than the primary key's term ID are considered secondary keys.
A count of 1 is added to each accumulator table associated with the primary keys detected in the current document.
The forward documents are then all scanned in turn to check for occurrences of the primary key; if a primary key is found, a count of 1 is added to each secondary key in the associated accumulator table.
Once all documents have been scanned, the counts are written out to disk and the process is repeated until each term in the vocabulary has had a turn as a primary key.
Multiple scans are necessary until all terms were processed as primary keys.
However, since terms in the forward documents are sorted by ascending term IDs and we only pair smaller IDed terms with those with a higher ID, the cost of scans is successively reduced.
After just a few passes many of the documents will have been fully processed and can be skipped entirely.
We found that using $a=100$ accumulators in parallel -- that is, a pass is performed for each 100 distinct terms in the collection -- achieves a reasonable balance between increased memory usage on the one hand (and the negative impact that this has on timing due to various factors), and fewer scans on the other.

\begin{table*}[t]
	\centering
	\small
		\begin{tabular}{lrrrrrrrrr}
                                   & 1     &    10 &    100 &  1000 &  10000 & 100000 & 1000000 & 1691666 \\
			\hline
Average document length            & 182.0 & 115.6 &  251.5 & 217.4 &  222.5 &  233.3 &   229.1 &   227.1 \\
Minimum document length            & 182   &    14 &     11 &     5 &      3 &      3 &       3 &       2 \\
Maximum document length            & 182   &   257 &   1171 & 2.19K &  8.57K &  47.8K &   73.6K &   73.6K \\
Document length standard deviation & 0.0   &  87.6 &    246 &   221 &    331 &    392 &     461 &     521 \\
Number of postings                 & 182   & 1.16K &  25.2K &  217K &  2.22M &  23.3M &    229M &    384M \\
Vocabulary size                    & 182   &   805 &  8.04K & 29.2K &   141K &   694K &   3.83M &   5.75M \\
Number of distinct Co-oc pairs     & 16.5K & 96.9K &  4.17M & 22.8M &   299M &  3.68B &   36.9B &   74.1B \\
Output size on disk                & 131KB & 767KB & 31.9MB & 174MB & 2.23GB & 27.4GB &   275GB &   552GB \\
			\hline
		\end{tabular}
	\caption{Collection statistics at various number of documents}\label{table:collectionStatistics}
\end{table*}

\section{Results}\label{section:results}
For our experiments we use the WT10G Web-document collection \cite{bailey2003engineering} with 1.69 million documents; properties are shown in Table~\ref{table:collectionStatistics}.
The collection is word-broken (using the Indri \cite{strohman2005indri} parser) and converted to lower case.
All per-document term repetitions are removed and finally the terms are replaced by 32-bit term identifiers.
Smaller collections are emulated by taking the first encountered documents in the test collection.
All experiments were run single-threaded under light load on a dual Intel Xeon E5-2643 CPU (with 8 cores) server running Windows 2012 (64-bit) with 64GB of RAM.

\begin{figure*}[t]
  \begin{minipage}[b]{1\textwidth}
	\includegraphics[width=\textwidth]{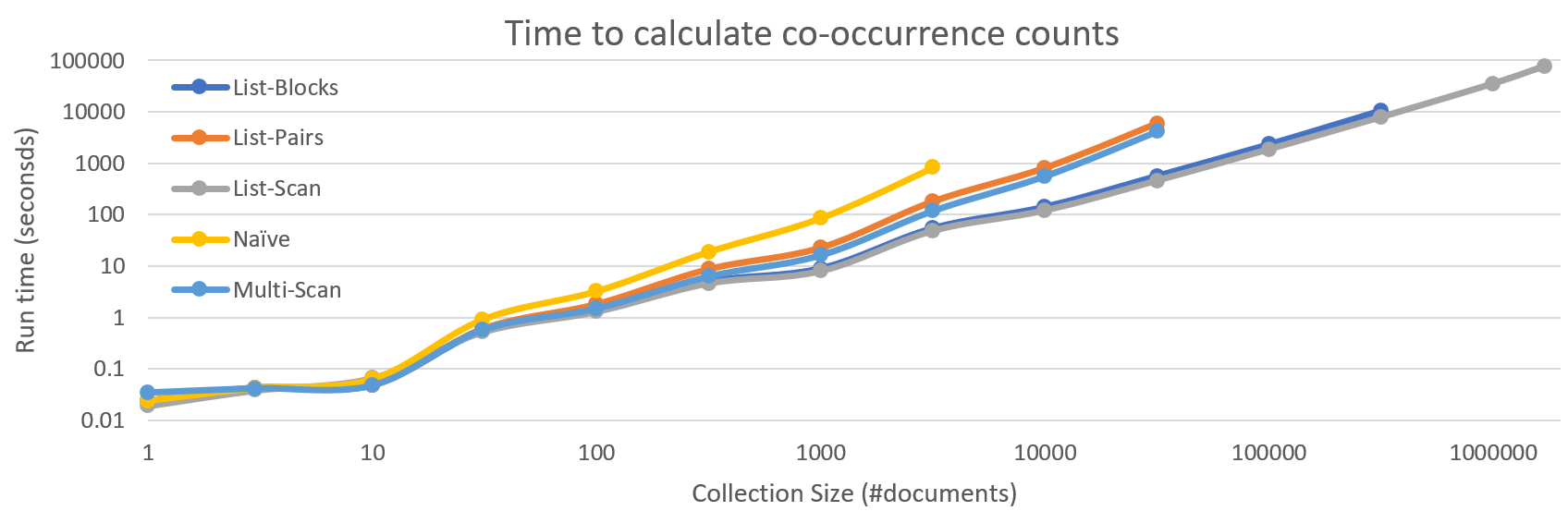}
	\caption{Time comparison of co-occurrence count methods}\label{figure:timings}
  \end{minipage}
\end{figure*}

\begin{figure*}[]
  \begin{minipage}[b]{1\textwidth}
	\includegraphics[width=\textwidth]{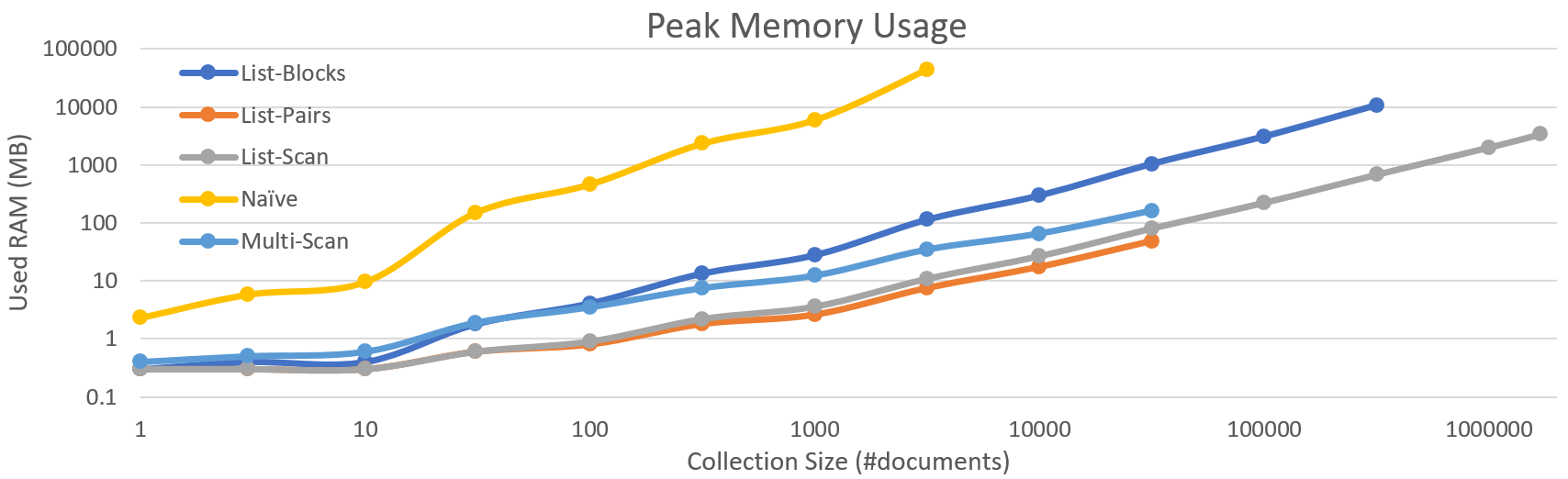}
	\caption{Memory usage of co-occurrence count methods}\label{figure:memoryUsage}
  \end{minipage}
\end{figure*}

Results are shown in Figures~\ref{figure:timings} and~\ref{figure:memoryUsage}.
These demonstrate substantial differences between the methods.
The \textsc{na\"ive} method in particular is both slow and memory-hungry,
requiring almost 2 hours (and at more than 200GB of RAM much swap space) for 10,000 documents;
it was not tested at larger collection sizes.
The \textsc{list-pairs} and \textsc{multi-scan} methods were also relatively slow, requiring over an hour for just over 30,000 documents, with time growing significantly worse than linearly in the number of documents.

However, the best two methods, \textsc{list-blocks} and \textsc{list-scan}, showed growth in time close to linear, and were able to process around 100,000 documents in about half an hour.
The timings for these two methods were remarkably similar, particularly given the very different algorithmic foundations.
To process the full collection -- almost 1.7M documents containing 5.7M distinct terms -- took just over 20 hours using the \textsc{list-scan} method, counting the frequency of 74 billion co-occurring term pairs.
Perhaps unsurprisingly, the most frequently co-occurring term pair was ``to''--``the'', which co-occurred in almost 1.3 million documents -- more than three-quarters of the document collection.

Several of these methods required that the collection be held in memory; for \textsc{list-blocks}, \textsc{list-scan}, and \textsc{list-pairs}, this was the dominant factor in memory consumption.
Scaling to larger collections would require that intermediate results be written and merged.
However, our results so far cannot be reliably extrapolated to this scenario.

Another factor impacting memory is the size of the largest document. 
In these experiments, we imposed no bounds on document length, and a single long document of 70,000 distinct terms (the largest in our collection) implies over two billion distinct term pairs.
A bound on document length would be required to cap either memory sizes or number of iterations.

For several of the methods, bounding of document length (say via windows) would also bound computation time.
Indeed, this approach is used in several of the methods that make use of term occurrence, often with very small windows of say 10 terms.
Our aim is to develop methods to work without such a bound, but a practical implementation for large document collections probably requires some restriction.

An open question is the performance of these methods on different kinds of collections. 
It is clear from the experiments so far that they respond in different ways to the different parameters (numbers of documents, document lengths) and thus further experiments are needed to confirm the generality of these methods.

\section{Discussion and Conclusions}\label{section:discussion}

Our experiments have shown that, with careful algorithmic design and implementation, accumulation of frequencies for all co-occurring term pairs is feasible with reasonable resources, on mo\-de\-ra\-te-sized collections.
In particular, the 
\textsc{list-blocks} and \textsc{list-scan} methods show manageable consumed memory and perform linearly in the size of the collection in our experiments.
However, care does need to be taken.
The nature of the task means that quadratic behaviour can emerge, leading to either large processing times or excessive memory requirements.
A careful complexity analysis should reveal under what conditions each algorithm performs quadratically, and may provide insight that leads to improved algorithms.
Similarly, analysis of memory and cache behaviour could lead to algorithms with increased performance.
Another interesting area for future study is examining how these algorithms scale to larger collections or operate in a distributed fashion.

Our most successful algorithms make combined use of inverted lists and forward indexes, allowing processing to proceed in pairwise order rather than document order and thus limiting requirements for memory and intermediate storage space.
However, their success also depends on careful implementation, and in particular we observed that preprocessing was critical to good performance.
We believe that further gains are possible within the framework of our existing algorithms, allowing processing of perhaps a million documents per hour on our hardware, but new algorithms will be required if web-scale collections are to be processed within reasonable resources.

\bibliographystyle{unsrt}

\bibliography{bibtex}
\end{document}